\documentclass[sigconf,screen]{acmart}

\AtBeginDocument{%
  }

\setcopyright{cc}
\setcctype{by}
\acmDOI{10.1145/3832783.3837543}
\acmYear{2026}
\copyrightyear{2026}
\acmISBN{979-8-4007-2882-2/2026/10}
\acmConference[ASE '26]{Proceedings of the 41st IEEE/ACM International Conference on Automated Software Engineering}{October 12--16, 2026}{Munich, Germany}
\acmBooktitle{Proceedings of the 41st IEEE/ACM International Conference on Automated Software Engineering (ASE '26), October 12--16, 2026, Munich, Germany}
\acmSubmissionID{ase26main-p2811-p}
\received{2026-03-26}
\received[accepted]{2026-06-18}

\ccsdesc[500]{Software and its engineering}

\usepackage{amsmath,amsfonts}
\usepackage{algorithmic}
\usepackage{graphicx}
\usepackage{balance}
\usepackage{textcomp}
\usepackage{listings}
\usepackage{xcolor}
\usepackage{hyperref}
\usepackage{url}
\usepackage{subcaption}
\usepackage[most]{tcolorbox}
\usepackage{enumitem}

\definecolor{codegreen}{rgb}{0,0.6,0}
\definecolor{codegray}{rgb}{0.5,0.5,0.5}
\definecolor{codepurple}{rgb}{0.58,0,0.82}
\definecolor{backcolour}{rgb}{1,1,1}

\newboolean{showcomments}
\setboolean{showcomments}{true}         
\ifthenelse{\boolean{showcomments}}
  {\newcommand{\nb}[2]{
  \fbox{\bfseries\sffamily\scriptsize#1}
     {\sf\small$\blacktriangleright$\textit{\textcolor{red}{#2}}$\blacktriangleleft$}
   }
  }
  {\newcommand{\nb}[2]{}
   
  }

\newcommand\new[1]{{\color{black}#1}}

\begin{document}

\title{When Knowledge Changes: Metamorphic Testing of RAG Systems with Mutations}

\author{Jinhan Kim}
\correspondingauthor
\orcid{0000-0002-0140-7908}
\affiliation{%
  \institution{Università della Svizzera italiana}
  \city{Lugano}
  \country{Switzerland}
}
\email{jinhan.kim@usi.ch}

\author{Samuele Pasini}
\orcid{0000-0002-7900-3727}
\affiliation{%
  \institution{Università della Svizzera italiana}
  \city{Lugano}
  \country{Switzerland}
}
\email{samuele.pasini@usi.ch}

\author{Paolo Tonella}
\orcid{0000-0003-3088-0339}
\affiliation{%
  \institution{Università della Svizzera italiana}
  \city{Lugano}
  \country{Switzerland}
}
\email{paolo.tonella@usi.ch}

\renewcommand{\shortauthors}{Kim et al.}

\begin{abstract}
Retrieval-Augmented Generation (RAG)-based LLM systems rely on external document corpora that can evolve and change over time. However, current evaluation methodologies (e.g., RAGAS) assess correctness against static snapshots, failing to detect faults when routine updates, factual changes, or noise alter the underlying data. We introduce a metamorphic testing framework that evaluates the consistency of RAG systems under corpus evolution. We formalise a fault taxonomy and 11 mutation operators that systematically perturb the system at both the pre-chunk (retrieval index) and post-chunk (retrieved context) levels. An empirical evaluation across five datasets and over 28k mutants reveals metamorphic violation rates of 4.9--10.2\%. In a meta-evaluation against ground truth, our metamorphic oracle achieves F1 scores of 0.927--1.000, while the best RAGAS metric reaches only 0.570. Finally, we provide actionable insights into mitigating these faults through retrieval re-configuration, generator upgrades, and LLM-based reranking.
\end{abstract}

\keywords{Retrieval-Augmented Generation, Metamorphic Testing}

\maketitle


\section{Introduction}

Retrieval-Augmented Generation (RAG) systems have become a dominant architectural pattern for deploying large language models (LLMs) in knowledge-intensive applications~\cite{lewis2020retrieval}. By retrieving documents from an external corpus and conditioning generation on the retrieved content, RAG systems aim to overcome the limitations of LLMs' parametric knowledge and provide answers grounded in external sources. This architecture is widely adopted in settings such as enterprise knowledge bases, compliance assistants, and proprietary documentation systems, where correctness depends on access to evolving document collections~\cite{ni2025towards}.

Existing evaluation methodologies for RAG systems primarily assess snapshot performance~\cite{es2024ragas, saadfalcon2023ares}. Benchmarks typically measure answer accuracy, retrieval quality, or grounding with respect to a fixed corpus representation. These evaluations are effective for comparing model configurations under controlled conditions, but they implicitly assume that the document corpus and retrieval pipeline remain unchanged. In practice, this assumption does not always hold: document corpora evolve continuously, facts are updated, rules change, evidence is added or removed, and documents are restructured, re-chunked, or re-indexed as part of system maintenance and evolution.

From a software engineering perspective, it is critical that a RAG system remains sensitive to the \textit{state} of the corpora. The system should not only provide correct answers for a static snapshot but also demonstrate appropriate behavioural shifts (or stability) as the corpus undergoes semantically significant (or insignificant) changes. For example, when an irrelevant document is re-formatted, the system’s output should remain unchanged; when a factual value is updated, the output should reflect the update; when supporting evidence is removed, the system should express uncertainty (e.g., abstain) rather than hallucinate. Whether current RAG systems satisfy such expectations remains unexplored.


To overcome this problem, we introduce a metamorphic testing paradigm specifically tailored for RAG systems. Rather than relying on a fixed ground-truth answer, we evaluate correctness by observing how system outputs shift in response to controlled semantic transformations of the corpus. We formalise these expectations as a set of Metamorphic Relations (MRs). These relations define the invariant properties of the system under corpus evolution, such as maintaining behavioural consistency when encountering irrelevant noise, updating answers appropriately when factual premises change, and explicitly acknowledging conflicts when contradictory evidence is injected.

Systematically driving this evaluation requires formalising how RAG systems fail. Based on a fault taxonomy that we derived by synthesising failure modes from prior empirical studies of RAG robustness~\cite{wu2025pandora, shi2023large, wu2024clasheval, liu2024lost, xu2023retrieval}, we design 11 mutation operators applied across two operational scopes: \textit{pre-chunk} mutations that perturb the upstream document index, and \textit{post-chunk} mutations that alter the downstream retrieved context directly. We then execute the RAG pipeline across the original and mutated context, checking the outputs against our defined MRs. Any violation serves as an indicator of semantic inconsistency, pointing to a potential weakness of the RAG system.

We evaluate our framework on five datasets, executing over 28k mutants across both scopes. Our study yields three principal findings. First, the framework detects faults across all five datasets, with violation rates ranging from 4.9\% to 10.2\%. The two mutation scopes expose complementary faults (Jaccard overlap 12--45\%), confirming the diagnostic value of testing at both the index and context levels. Second, in a meta-evaluation against ground-truth correctness, our MR oracle achieves F1 scores of 0.927--1.000, whereas the best-performing RAGAS metric~\cite{es2024ragas} reaches only 0.570 F1.
\new{Third, repair effectiveness varies substantially across datasets. The union of retrieval-budget increases, a generator upgrade, and LLM-based reranking resolves 43.1\% of evaluated faults. Union repair rates range from 47.3\% to 63.0\% across the four financial datasets but reach only 10.2\% on RepLiQA, pointing to failure modes that neither broader retrieval, stronger generation, nor LLM-based reranking can fully address.}


In summary, this paper makes the following contributions:

\begin{itemize}
    \item We introduce a metamorphic testing framework for RAG systems under corpus evolution, driven by 11 mutation operators across two scopes.

    \item \new{We conduct an empirical study across five datasets demonstrating that our MRs expose silent failures that RAGAS metrics overlook, validated by a human study and oracle stability analysis.}

    \item \new{We analyse the repairability of detected faults, showing that LLM-based reranking complements retrieval and generator interventions and that their union repairs 43.1\% of detected faults.}
    

\end{itemize}


\section{Background}
\label{sec:background}

This section summarises established concepts and practices relevant to this work.

\subsection{Retrieval-Augmented Generation (RAG)}
\label{sec:background_rag}

RAG refers to a class of systems that combine a language model with an external document collection~\cite{gan2025retrieval}. Given a user query, a retrieval component selects a set of documents from the corpus, and a generation component (e.g., LLM) produces an output conditioned on both the query and the retrieved documents. RAG systems are commonly used when the information required to answer queries is not fully captured in the model’s parameters, or when access to external, domain-specific, or up-to-date knowledge is required. The document corpus is therefore treated as an integral part of the system rather than as static training data.

The standard RAG procedure consists of two primary phases: document indexing and output generation. During the offline indexing phase, the corpus is prepared by parsing and segmenting documents into smaller textual units called \textit{chunks}. These chunks are transformed into high-dimensional vectors via an embedding model~\cite{devlin2019bert} and stored in a vector database. This spatial arrangement is fundamental to the system's behaviour, as proximity within the vector space indicates semantic similarity between pieces of information.
In the subsequent online inference phase, the system processes a user query by embedding it with the same model to identify and retrieve the \textit{top-k} most similar chunks using a distance measure like cosine similarity. These chunks serve as the essential context for the system’s decision-making and are integrated into a structured prompt. The process concludes when the LLM generates a response that is specifically conditioned on both the original query and this retrieved evidence.

\subsection{Evaluation of RAG Systems}
\label{sec:background_rag_eval}


Frameworks like RAGAS~\cite{es2024ragas} and ARES~\cite{saadfalcon2023ares} propose automated, reference-free evaluation metrics that approximate human judgements using LLM-as-a-judge. For example, at the retrieval level, \textit{Context Precision} calculates the fraction of retrieved chunks that are genuinely relevant to the query. Similarly, \textit{Context Relevance} evaluates the context at the sentence level by measuring the ratio of sentences strictly necessary to answer the query against the total number of sentences retrieved.
At the generation level, \textit{Response Relevancy} estimates how well the generated answer addresses the original query. It achieves this by prompting an auxiliary LLM to generate new questions based on the generated answer and calculating their semantic similarity to the original query. Finally, \textit{Faithfulness} evaluates whether the answer is actively supported by the retrieved context. It decomposes the response into atomic statements and calculates the proportion of these statements backed by the retrieved chunks, which is crucial for identifying hallucinated claims.

While these metrics are useful for comparing systems under controlled conditions, they provide limited insight into testing how a RAG system behaves as its underlying knowledge base evolves. This highlights the need for evaluation frameworks that assess semantic consistency upon knowledge evolution, ensuring the system remains robust under realistic corpus perturbations.


\subsection{Metamorphic Testing}
\label{sec:background_mr}

Metamorphic Testing (MT) is a testing methodology developed for systems where test oracles are unavailable, incomplete, or impractical~\cite{segura2016survey}. Instead of checking individual outputs for correctness, MT specifies relations that should hold between outputs produced under related inputs or environments. Violations of Metamorphic Relations (MRs) indicate unexpected or inconsistent system behaviour, even when individual outputs may appear plausible.

In this work, MT provides a basis for evaluating RAG systems by comparing system behaviour across different corpus states without assuming a fixed oracle. Instead of checking whether an output is absolutely correct against a static ground truth, we define a set of MRs that specify how the system's outputs should relate when the document corpus undergoes controlled semantic changes. This relational approach allows us to define correctness as the satisfaction of specific predicates across original and mutated corpora. By checking these relationships, we expose structural faults and reasoning errors that may remain undetected by standard evaluation methods.

\section{System Model and Fault Taxonomy}
\label{sec:system_model_and_fault_taxonomy}

In this section, we first formalise the RAG architecture as a composite pipeline of retrieval and generation functions governed by a set of system parameters (Section~\ref{sec:system_model}). We then introduce a fault taxonomy that classifies robustness failures based on the system's inability to maintain semantic consistency under controlled perturbations to its corpus or configuration (Section~\ref{sec:fault_taxonomy}).

\begin{figure*}[t]
    \centering
    \includegraphics[width=0.73\linewidth]{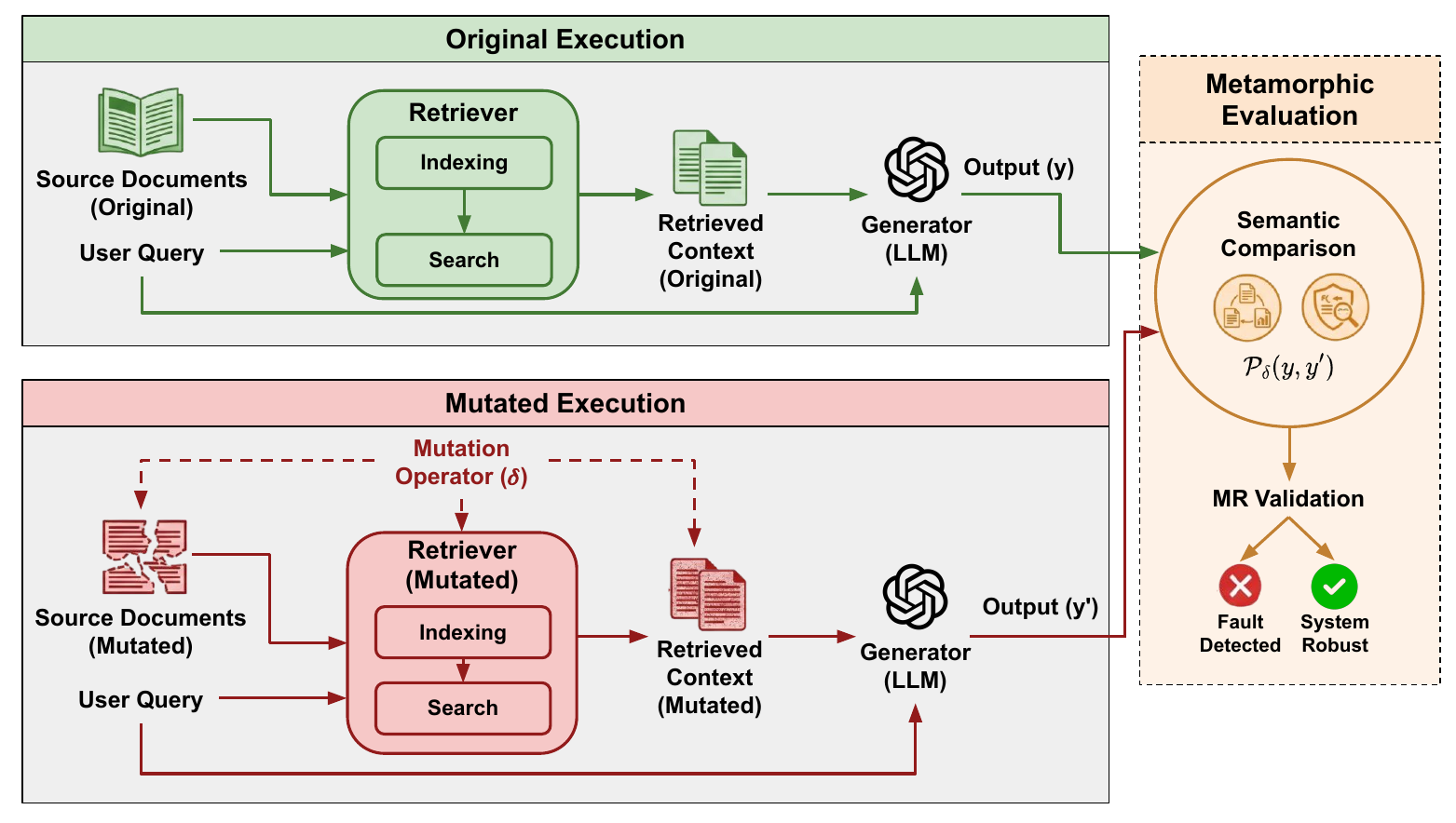}
    \caption{Overview of our metamorphic testing framework for RAG systems. The original execution (top) establishes a baseline by processing source documents and a user query through standard indexing and search to produce an original answer ($y$). The mutated execution (bottom) introduces a mutation operator ($\delta$) that perturbs the system. The metamorphic evaluation component performs a semantic comparison between two outputs ($y$ and $y'$). A violation of the metamorphic relation ($\neg \mathcal{P}_\delta$) indicates a detected fault, whereas semantic consistency suggests system robustness.}
    \label{fig:overview}
\end{figure*}

\subsection{System Model}
\label{sec:system_model}

The overview of our approach is depicted in Figure~\ref{fig:overview}. We treat the RAG pipeline as a stateful process where the transition from input query to output answer is mediated by the retrieval of evidence from the corpus. By introducing systematic perturbations to the source documents, retrieved context, or system settings, we would (or would not) observe the resulting shifts in the output. This differential execution allows us to isolate the causal chain of a failure: a fault is not merely a wrong answer, but a deviation from the expected semantic relationship between two executions.

We formalise a RAG system $\mathcal{S}$ as a composite function of a retriever $\mathcal{R}$ and a generator $\mathcal{G}$, operating over a dynamic corpus $\mathcal{C}$. To capture the highly configurable nature of these pipelines, we denote the overall system configuration as $\Theta = \langle \theta_{\mathcal{R}}, \theta_{\mathcal{G}} \rangle$, where $\theta_{\mathcal{R}}$ represents retrieval hyperparameters (e.g., top-$k$, chunk size, or embedding model) and $\theta_{\mathcal{G}}$ represents generation hyperparameters (e.g., LLM temperature, LLM model version).\footnote{We distinguish the generator parameters ($\theta_{\mathcal{G}}$) from the retrieval parameters ($\theta_{\mathcal{R}}$) because $\theta_{\mathcal{G}}$ could theoretically be mutated to expose faults. However, our proposed mutation operators do not alter $\theta_{\mathcal{G}}$ because this work primarily targets the retrieval component of the RAG pipeline. Nevertheless, one of our research questions (RQ3) explores adjusting $\theta_{\mathcal{G}}$ to \textit{fix} a fault once it has been localised to the generator rather than the retriever (see Section~\ref{sec:results_rq3}).} Given a query $q$, the system produces an output
$y = \mathcal{S}_{\Theta}(q, \mathcal{C}) = \mathcal{G}_{\theta_{\mathcal{G}}}\big(q, \mathcal{R}_{\theta_{\mathcal{R}}}(q, \mathcal{C})\big)$

In real-world deployments, neither $\mathcal{C}$ nor $\Theta$ is static. The corpus evolves through updates, segmentation changes, or noise accumulation, while system configurations are frequently tuned to optimise performance. This environmental and configurational volatility renders traditional test oracles (fixed $\langle q, a \rangle$ pairs) insufficient, as the correct answer $a$ may change or become invalid under different system states.

To address this oracle problem, we adopt MT. We define correctness not as an absolute match against a static ground truth, but as the satisfaction of MR. Let $\delta$ be a mutation operator that perturbs the system state such that $\langle \mathcal{C}', \Theta' \rangle = \delta(\mathcal{C}, \Theta)$. A robust RAG system must satisfy a specific metamorphic relation $\mathcal{P}_\delta$ between the original output $y$ and the post-mutation output $y'$:
$$\forall q, \mathcal{C}, \Theta: \quad \mathcal{P}_\delta\big(\mathcal{S}_{\Theta}(q, \mathcal{C}), \mathcal{S}_{\Theta'}(q, \mathcal{C}')\big)$$

In this framework, a \textit{fault} is defined as a violation of this predicate ($\neg \mathcal{P}_\delta$). For example, if $\delta$ injects irrelevant noise into the corpus, $\mathcal{P}_\delta$ requires semantic invariance ($y \approx y'$). Conversely, if $\delta$ updates a fact in the corpus or severely restricts $\theta_{\mathcal{R}}$, causing essential chunks to drop, $\mathcal{P}_\delta$ dictates a corresponding output update or a fallback response. The following taxonomy classifies these violations based on the semantic nature of the failure.

\subsection{Fault Taxonomy}
\label{sec:fault_taxonomy}

To systematically characterise failures in RAG systems, we define a fault model based on the system's inability to maintain semantic consistency under perturbations. 
Our taxonomy was constructed by synthesising failure modes reported in prior empirical studies of RAG robustness. We began from the RAG noise taxonomy of Wu et al.~\cite{wu2025pandora}, which identifies noise types that degrade LLM performance, and extended it with findings from studies on irrelevant context sensitivity~\cite{shi2023large}, knowledge conflicts between retrieved and parametric knowledge~\cite{wu2024clasheval}, positional bias in long contexts~\cite{liu2024lost}, and retrieval configuration sensitivity~\cite{xu2023retrieval}. We consolidated these into three fault classes: (1) \textit{Noise Sensitivity Faults}, extending the RAG noise taxonomy established by Wu et al.~\cite{wu2025pandora}; (2) \textit{Format Intolerance Faults}; and (3) \textit{Architectural Brittleness Faults}, a classification addressing structural brittleness in retrieval pipelines. This taxonomy serves as the foundation for the mutation operators defined in Section~\ref{sec:operators}.

\subsubsection{Noise Sensitivity Faults}
This category encompasses faults where the generator (i.e., LLM) fails to correctly filter or arbitrate external information, leading to output corruption.

\begin{itemize}[leftmargin=0.3cm]
    \item \textit{Context Overfitting Fault}: A logic fault where the system treats topically related but information-poor padding as valid evidence. In production environments, this manifests when retrieval systems surface content optimised for search engines or advertisements that share keywords with the query but lack factual substance. Shi et al.~\cite{shi2023large} demonstrate that LLMs are highly susceptible to such irrelevant context, often prioritising retrieved distractors over relevant knowledge.
    
    \item \textit{Knowledge Conflict Fault:} A state inconsistency fault where the system fails to resolve contradictory facts. This reflects the common real-world challenge of \textit{stale data}, where a knowledge base contains both outdated legacy documents and recent updates (e.g., v1 vs. v2 manuals). Wu et al.~\cite{wu2024clasheval} demonstrate that systems frequently exhibit `context bias', overriding correct internal priors to align with contradictory retrieved evidence. A robust system must exhibit explicit conflict awareness, signalling or resolving the discrepancy.
    
    \item \textit{Input Robustness Fault}: A failure to handle minor character-level perturbations, resulting in significant degradation of retrieval precision or generation quality~\cite{wu2025pandora}. This fault is critical in pipelines processing OCR-scanned legacy documents, noisy speech-to-text transcripts, or user-generated content containing informal spelling.
\end{itemize}

\subsubsection{Format Intolerance Faults}
Although recent studies suggested that mixing data types should theoretically help models distinguish between plain text and core data, many RAG systems still fail under these conditions~\cite{wu2025pandora}. Such faults typically occur because the system lacks the flexible parsing logic required to process unexpected structural changes.

\begin{itemize}[leftmargin=0.3cm]
    \item \textit{Heterogeneous Data Fault}: The system can fail or hallucinate when processing mixed-modality content. This is a pervasive issue in technical domains where documentation frequently interleaves natural language with code snippets (Python, SQL), JSON configurations, or tabular financial data.
    \item \textit{Syntactic Filtering Fault}: The system attempts to reason over unintelligible, garbled token sequences instead of discarding them as noise. This simulates the faults in data cleaning pipelines, where HTML artefacts, encoding errors (mojibake), or truncated data packets pollute the retrieval index, leading to garbage-in-garbage-out behaviour.
\end{itemize}

\subsubsection{Architectural Brittleness Faults}
This category describes faults where the system's correctness is tightly coupled to specific implementation details or parameters of the retrieval pipeline ($\theta_{\mathcal{R}}$), rather than the semantic content of the knowledge base. A robust RAG system should exhibit \textit{operational invariance}, producing consistent outputs regardless of choices in data discretisation or retrieval configuration.\footnote{However, note that we scope this expectation to non-disruptive changes; radical configuration changes that fundamentally alter the retrieved evidence would justifiably produce different outputs (see Section~\ref{sec:metamorphic_relationships_as_test_oracles} for details).}

\begin{itemize}[leftmargin=0.3cm]
    \item \textit{Segmentation Boundary Fault:} A fault where the system's output is unstable with respect to the specific discretisation (chunking) of the source text. In practice, fixed-size windowing often arbitrarily severs semantic dependencies, splitting a disclaimer from its clause or a subject from its verb. This sensitivity is empirically linked to the ``lost-in-the-middle'' phenomenon identified by Liu et al.~\cite{liu2024lost}, where models fail to retrieve relevant information solely because the chunking strategy shifted the evidence away from the edges of the context window.
    
    \item \textit{Configuration Overfitting Fault:} This represents a dependency fault where correctness is strictly tied to specific chunk alignments within the index or a fixed retrieval budget defined in $\theta_{\mathcal{R}}$. It demonstrates that the system is brittle and overfitted to a narrow range of configurations. Xu et al.~\cite{xu2023retrieval} highlight that architectural choices in retrieval depth can unexpectedly degrade performance, exposing this underlying lack of generalisation.
\end{itemize}

\section{Mutation Operators and Test Oracles}
\label{sec:operators}

We define a set of 11 mutation operators to realise the fault taxonomy defined in Section~\ref{sec:fault_taxonomy}.
In the following sub-sections, after defining the scope of mutation, we introduce our novel mutation operators, followed by a description of the metamorphic relations used as oracles.

\subsection{Mutation Scope}
\label{sec:mutation_scope}

We evaluate RAG robustness across two distinct operational stages, pre-chunking and post-chunking. By perturbing the system at different stages, we can assess how changes to the knowledge base and the retrieved context independently influence the reliability of the generated output.

\subsubsection{Pre-chunk Scope (System-Level Mutation)}

This scope performs system-level mutation by modifying the raw documents or retrieval configurations \textit{upstream} of the ingestion pipeline. Formally, we define the mutated system state as $\langle \mathcal{C}', \theta_{\mathcal{R}}' \rangle = \delta_{pre}(\mathcal{C}, \theta_{\mathcal{R}})$, resulting in a perturbed output $y' = \mathcal{G}_{\theta_{\mathcal{G}}}(q, \mathcal{R}_{\theta_{\mathcal{R}}'}(q, \mathcal{C}'))$. This perturbation alters the documents in the underlying corpus or the retrieval parameters (such as embedding models or chunking strategies), requiring a full re-indexing of the knowledge base. Unlike simple input fuzzing for LLMs, this modifies the index structure and knowledge representation, simulating the operational reality of continuous document edits, policy updates, or architectural migrations.

\subsubsection{Post-chunk Scope (Input-Level Mutation)}

This scope performs input-level mutation (analogous to standard input fuzzing) targeting the retrieved context passed from the retriever to the generator. While less realistic, it can provide a more efficient alternative to pre-chunk mutation as it bypasses the re-indexing of the corpus or the re-computation of embeddings. Formally, we define the perturbed output as $y' = \mathcal{G}_{\theta_{\mathcal{G}}}\big(q, \delta_{post}(\mathcal{R}_{\theta_{\mathcal{R}}}(q, \mathcal{C}))\big)$, where $\delta_{post}$ perturbs the retrieved context before it reaches the generator. Mutations are injected \textit{downstream} of the retrieval step, modifying only the transient prompt while keeping the retrieval index $\mathcal{C}$ and configurations $\theta_{\mathcal{R}}$ and $\theta_{\mathcal{G}}$ pristine. This approach serves as a controlled experiment: by guaranteeing the mutation is present in the prompt, we isolate the generator and stress-test its reasoning capabilities independently of any retrieval variance. Consequently, this scope simplifies fault localisation; if a failure occurs under post-chunk mutations, the behavioural change can be attributed to the generator.

\subsection{Noise Mutation Operators}
\label{sec:noise_mutation_op}

These operators inject textual perturbations to expose \textit{Noise Sensitivity} and \textit{Format Intolerance} faults.

\textbf{Semantic Noise Operator (SeN).} SeN targets \textit{Context Overfitting Faults} by injecting off-topic sentences from a static pool of unrelated facts. Pre-chunk injection scatters distractors throughout the document, shifting chunk boundaries and polluting the index to test if the retriever maintains precision against irrelevant signals. Post-chunk injection appends off-topic text to retrieved chunks, testing the LLM's ability to filter irrelevant trailing context.

\textbf{Datatype Noise Operator (DN).} DN targets \textit{Heterogeneous Data Faults} by injecting foreign fragments (code, URLs, JSON) into prose. Fragments are drawn from a predefined pool of code snippets and inserted at sentence boundaries. Pre-chunk injection embeds these into the corpus structure, testing vector space robustness against mixed modalities. Post-chunk injection injects fragments into retrieved chunks individually, validating if the LLM can parse mixed-type inputs without hallucination.

\textbf{Illegal Sentence Noise Operator (ISN).} ISN targets \textit{Syntactic Filtering Faults} by inserting syntactically malformed `garbage' sentences constructed by randomly sampling and concatenating words from a predefined vocabulary of common English tokens (e.g., articles, verbs, nouns), producing syntactically incoherent text. Pre-chunk injection distributes this noise across the document, potentially causing the retriever to index unintelligible chunks. Post-chunk injection appends garbled text to retrieved chunks, testing the LLM's exception handling and garbage-in-garbage-out resilience.

\textbf{Counterfactual Noise Operator (CN).} CN targets \textit{Knowledge Conflict Faults} by injecting statements that contradict ground truth. To produce realistic contradictions, the operator prompts a separate LLM to generate a plausible counterfactual statement given the original fact with the user query. In pre-chunk injection, contradictions are embedded in the corpus; critically, the retriever may return the contradiction without the original fact, simulating stale data scenarios. In post-chunk injection, a contradiction is generated for and appended to the retrieved chunks, forcing the LLM to arbitrate immediate logical conflicts between parametric priors and external evidence.

\textbf{Supportive Noise Operator (SuN).} SuN targets \textit{Context Overfitting Faults} by adding topically related but information-void fillers. The operator heuristically extracts key terms (e.g., domain nouns) from the source text and inserts them into predefined sentence templates, producing filler that sounds on-topic but contains no specific facts. Pre-chunk injection dilutes the corpus, challenging dense retrievers that may rank filler-heavy chunks highly due to semantic overlap. Post-chunk injection appends fillers to retrieved chunks, testing if the LLM is distracted by relevant-sounding padding.

\textbf{Orthographic Noise Operator (ON).} ON targets \textit{Input Robustness Faults} by applying character-level typos. Pre-chunk injection degrades the index, testing fuzzy matching between clean queries and misspelled documents. Post-chunk injection preserves retrieval but forces the LLM to reason over degraded surface forms.

\subsection{Chunking Mutation Operators}
\label{sec:chunking_mutation_op}

These operators alter the segmentation strategy to expose \textit{Segmentation Boundary Faults}.

\textbf{Chunk Merge (CM) \& Chunk Split (CSp) Operators.} CM aggregates adjacent chunks into fewer, larger index entries, altering the granularity of the retrieval index. CM is only applicable in pre-chunk mode; at the post-chunk stage, the retriever has already returned individual chunks, so there are no adjacent chunks to merge. Conversely, CSp subdivides chunks. Pre-chunk injection fragments the index, increasing the risk of relevant content falling below retrieval thresholds; post-chunk injection forces the LLM to synthesise information across a highly fragmented context list.

\textbf{Chunk Shift Operator (CSh).} CSh re-chunks the document starting from an offset (e.g., 50\%). Pre-chunk injection performs global re-chunking, creating entirely new boundary alignments across the corpus. Post-chunk injection performs a local token shift, trimming tokens from the start of a retrieved chunk and prepending them to the next. This operator directly provokes the ``lost-in-the-middle'' phenomenon~\cite{liu2024lost} by arbitrarily shifting the position of key evidence within the context window.


\subsection{RAG Configuration Operators}
\label{sec:config_mutation_op}

These operators perturb the execution environment to evaluate the system against \textit{Configuration Overfitting Faults}. By directly modifying the retrieval hyperparameters $\theta_{\mathcal{R}}$ rather than the corpus text, these operators are exclusively pre-chunk: they require re-indexing or re-retrieval, which has no analogue at the post-chunk stage where the retrieved context is already fixed.

\textbf{Embedding Model Swap (EM).} EM replaces the retrieval hyperparameter in $\theta_{\mathcal{R}}$ (e.g., swapping \texttt{text-embedding-3-small} for \texttt{large}) to evaluate whether system performance is coupled to a specific vector space. A robust RAG implementation should exhibit operational invariance across different embedding geometries.

\textbf{Top-$k$ Perturbation (TK).} \new{TK modifies the retrieval hyperparameter in $\theta_{\mathcal{R}}$ by altering the number of chunks passed to the generator. Restricting $k$ evaluates whether the generator can maintain precision when relying on fewer retrieved chunks, whereas expanding $k$ tests whether it can isolate relevant evidence from redundant or distracting context. However, beyond its role as a mutation operator, we note that changing $k$ may also \textit{repair} a fault. As a repair operator, increasing $k$ may recover missing evidence, while decreasing $k$ may filter out distractions. The same type of top-$k$ change can therefore induce or repair a fault depending on its context, consistent with the inversive relationship between bugs and patches observed by Kim et al.~\cite{kim2023inversive}. This motivates our evaluation of retrieval-budget changes as a repair intervention in RQ3 (see Section~\ref{sec:results_rq3}).}



\subsection{Metamorphic Relationships as Test Oracles}
\label{sec:metamorphic_relationships_as_test_oracles}

We define MRs that verify the logical consistency between RAG system executions. Specifically, we implement the metamorphic predicate $\mathcal{P}_\delta$ (defined in Section~\ref{sec:system_model}) to enforce principles of \textit{semantic invariance} or \textit{controlled variance} across mutations. Our oracles are categorised into two classes based on the expected transformation of the system's response:

\begin{itemize}[leftmargin=0.5cm]
    \item \textit{Invariance Relations (Semantic-Preserving):} For a mutation operator $\delta_{p}$ that preserves the semantic essence of the corpus to answer the query, a robust system must exhibit operational invariance. We expect $y \approx y'$, where $\approx$ denotes semantic equivalence. 
    
    \item \textit{Variance Relations (Semantic-Altering):} For a mutation operator $\delta_{a}$, e.g., that injects counterfactual facts related to the query, the system must update its output to reflect the new state of the knowledge base. We expect $y' \neq y$, specifically where $y'$ aligns with the injected fact. 
\end{itemize}

It is important to note that, in reality, the classification of an MR as semantic-preserving or semantic-altering is a function of both the mutation operator type and its specific parameters. For example, with the ISN operator, a 5\% noise injection ratio is intended to be semantic-preserving, whereas a 95\% injection (replacing most content with garbled text) is likely semantic-altering. Evaluating both the mutation's effect and the system's response requires judgements that go beyond simple string matching. We employ an LLM-as-a-Judge~\cite{zheng2023judging} (detailed in Section~\ref{sec:llm-as-a-judge}) in two roles: (1) a \textit{semantic preservation judge} that verifies, for each mutant, whether the factual evidence required to answer the query remains intact after the perturbation, determining the actual semantics-preserving or -altering status rather than relying on the operator's intended category; and (2) an \textit{answer equivalence judge} that assesses whether $y$ and $y'$ are semantically equivalent given the query and context, implementing the $\approx$ / $\neq$ relations above.

When the RAG system \textit{abstains}, the verdict is assigned deterministically without invoking the LLM judge. For semantics-preserving mutations, abstention constitutes an MR violation; for semantics-altering mutations, abstention is treated as MR satisfaction.


\section{Experimental Setup}
\label{sec:experimental_setup}

\subsection{Research Questions}
\label{sec:research_questions}

We structure our empirical study around the following three research questions:

\noindent \textbf{RQ1 (Fault Detection):} \textit{Can our metamorphic framework detect  RAG systems' faults?}
We quantify the system's susceptibility to noise and structural perturbations by investigating the rates of MR violations per mutation operator. Additionally, we conduct a comparative analysis between pre-chunk (system-level) and post-chunk (input-level) mutations to assess their respective effectiveness and identify the overlap in the faults detected.

Lastly, we conduct a human validation study to evaluate the reliability of our automated oracle. By measuring the alignment between human annotators and the LLM-as-a-Judge regarding both the semantic preservation of the mutation and the equivalence of the generated answers, we aim to prove that our automated evaluation framework is trustworthy.

\noindent \textbf{RQ2 (Comparison):} \textit{Do our MRs reveal the faults that existing RAG evaluation metrics overlook?}
\new{We compare our MRs with four representative snapshot metrics from RAGAS. This comparison does not treat RAGAS as a robustness oracle: RAGAS evaluates an individual answer and its retrieved context, whereas our MRs evaluate the relationship between two executions before and after a controlled mutation. We therefore examine whether snapshot metrics detect the same evolution-induced faults exposed by relational testing. Using ground-truth correctness as the meta-evaluation oracle, we compute precision, recall, and F1-score for each evaluator and analyse their fault overlap using Venn diagrams. We also assess the stability of the MR oracle on RepLiQA, where answer-equivalence judgements require LLM-based evaluation. We stratify-sample 100 cases by mutation scope, operator, initial verdict, and semantic-preservation label, then repeat the final MR judge verdict five times per case.}



\noindent \textbf{RQ3 (Repair):} \new{\textit{Can identified faults be mitigated by retrieval, generation, or reranking interventions?}}
\new{We investigate the persistence of identified faults under three repair interventions: increasing the retrieval budget, upgrading the generator model, and inserting an LLM-based reranking stage before generation. For pre-chunk faults, we evaluate direct retrieval with $k=3$ where applicable and with $k=5$, together with the reranking. For post-chunk faults, we evaluate an upgrade from GPT-5-mini to GPT-5.4 and reranking over the mutated retrieved context. The reranking strategy partitions its candidates into at most 30 evidence units, uses an LLM to score each question-unit pair from 0 to 5, and selects up to six of the highest-ranked units for answer regeneration.}

\subsection{Experimental Procedure}
\label{sec:experimental_procedure}

Our evaluation employs a multi-stage pipeline designed to isolate the causal impact of mutations on RAG behaviour. We utilise datasets (detailed in Section~\ref{sec:datasets}) comprising a query $q$, a corpus $\mathcal{C}$ (documents or relevant information), and a GT answer $a_{gt}$. While a primary advantage of our metamorphic approach is its independence from GT labels in production, we leverage $a_{gt}$ in the study to facilitate a rigorous meta-evaluation of \textit{evaluators}. 

\subsubsection{Filtering for Context Dependency} \label{sec:filtering}
To ensure our test suite targets the retrieval pipeline rather than the generator's parametric memory, we apply a \textit{closed-book filter}. We first execute the generator in isolation (without $\mathcal{C}$): $y_{closed} = \mathcal{G}(q)$. We retain only those triplets $\langle q, \mathcal{C}, a_{gt} \rangle$ in the dataset where the closed-book response $y_{closed}$ is \textit{incorrect}, i.e., $y_{closed} \neq a_{gt}$, while with $\mathcal{C}$, i.e., generating the answer using the full RAG pipeline, $y_{base} = \mathcal{S}(q, \mathcal{C})$ successfully produces the ground-truth answer $a_{gt}$, i.e., $y_{base} \approx a_{gt}$. This step ensures that the retrieved evidence is a \textit{necessary condition} for correctness, allowing us to attribute subsequent failures to the perturbation $\delta$.

\subsubsection{Metamorphic Execution}
For each filtered instance, we treat the successful baseline output $y_{base}$ as our reference point ($y$ in Figure~\ref{fig:overview}). We then apply a mutation $\delta$ to generate a post-mutation output $y'$.
Unlike traditional benchmarks that compare $y'$ to a static ground truth $a_{gt}$, our metamorphic oracle evaluates the logical consistency between $y_{base}$ and $y'$. We say a fault is detected by our framework if the answer pair $\langle y_{base}, y' \rangle$ violates the metamorphic predicate $\mathcal{P}_\delta$.

\subsubsection{Use of LLM-as-a-Judge}
\label{sec:llm-as-a-judge}

The complexity of evaluating answer pairs (e.g., comparing $a_{gt}$ with $y_{base}$, or $y_{base}$ with $y'$) depends heavily on the output format. For numerical outputs in our financial dataset, T2-RAGBench (see Section~\ref{sec:datasets}), we perform direct comparison to determine correctness. However, when the generator produces natural language outputs, assessing logical equivalence becomes non-trivial.

To address this, we employ an LLM-as-a-Judge~\cite{zheng2023judging}. The judge is provided with the query $q$, the retrieved contexts, information about the perturbation $\delta$, and the respective answer pairs to determine their relationship under $\mathcal{P}_\delta$. In addition to evaluating the system's output, we utilise the LLM-as-a-Judge to verify the nature of the mutation itself. Specifically, the judge determines whether a perturbation is semantic-preserving (where the factual evidence required to answer $q$ remains intact) or semantic-altering. This automated verification is critical because the disruptiveness of the mutation operators is highly sensitive to their parameterisation. For instance, operators such as SeN can vary from subtle stylistic changes to highly disruptive perturbations if a high volume of noise is injected into the document. 

Although the final verdict of the judge is binary, we utilise Likert scoring-based prompting to capture logical and linguistic nuances, a method shown to be more effective than simple prompting~\cite{wang2025improving}. For example, in the case of semantic-altering mutations, the judge evaluates how effectively the system updates its response to reflect new or counterfactual information on a 1--5 scale. High scores of 4--5 indicate successful adaptation, where $y'$ clearly reflects the mutated context or appropriately acknowledges conflicting information, satisfying the $\mathcal{P}_\delta$ relation. Conversely, scores of 1--2 indicate a failure to adapt, where $y'$ remains essentially identical to $y_{base}$ despite the change in evidence, resulting in an MR violation.

We acknowledge that relying on the LLM-as-a-Judge introduces a potential threat to validity due to hallucinations of the judges.\new{To assess this threat, we conduct two complementary analyses. First, two authors independently annotated a random sample of 100 RepLiQA mutation pairs, for both semantic preservation and answer equivalence. Second, we assessed oracle stability using 100 RepLiQA cases stratified by mutation scope, operator, initial verdict, and semantic-preservation label, repeating the final MR judgement five times for each case. Section~\ref{sec:results_rq1} reports the human agreement and oracle stability results.}


\subsection{Meta-Evaluation for Enabling Comparison}
\label{sec:meta_evaluation}

To quantify the diagnostic accuracy of our framework, we introduce a meta-evaluation procedure that assesses the effectiveness of our framework against established RAG evaluation metrics, response relevancy, faithfulness, context precision, and context recall, provided by the RAGAS framework~\cite{es2024ragas}. This stage effectively serves to evaluate the evaluators by measuring how reliably each oracle detects system failures. While a core advantage of MR is its ability to operate without a ground-truth oracle, we selectively re-introduce $a_{gt}$ here as an objective baseline to determine absolute factual correctness under mutation. With GT, we map the verdicts of both RAGAS and our MR framework into a formal classification of True Positives (TP), False Positives (FP), True Negatives (TN), and False Negatives (FN), and corresponding recall, precision, and F1-score.

For our framework, we compare the metamorphic verdict ($\mathcal{P}_\delta$) against the logical relationship between the post-mutation output ($y'$) and the ground truth ($a_{gt}$). For the RAGAS metrics, we define a fault detection as any score falling below a set threshold. This mapping allows us to identify \textit{silent faults}. For instance, if a semantic-preserving mutation results in an output $y'$ that contradicts $a_{gt}$ while a RAGAS metric maintains a high score, the metric suffers an FN, failing to register a structural fragility that our MR framework is designed to expose.


\begin{table*}[t]
\centering
\small
\caption{Pre-chunk and Post-chunk MR Violation Rates (\%) with Fault Overlap (Jaccard \%)}
\label{tab:rq1_violations_overlap}

\begin{tabular}{l|rrr|rrr|rrr|rrr|rrr}
\toprule
 & \multicolumn{3}{c|}{\textbf{RepLiQA}} & \multicolumn{3}{c|}{\textbf{ConvFinQA}} & \multicolumn{3}{c|}{\textbf{FinQA}} & \multicolumn{3}{c|}{\textbf{TAT-DQA}} & \multicolumn{3}{c}{\textbf{VQAonBD}} \\
\cmidrule(lr){2-4} \cmidrule(lr){5-7} \cmidrule(lr){8-10} \cmidrule(lr){11-13} \cmidrule(lr){14-16}
\textbf{MO} & \textbf{Pre} & \textbf{Post} & \textbf{J} & \textbf{Pre} & \textbf{Post} & \textbf{J} & \textbf{Pre} & \textbf{Post} & \textbf{J} & \textbf{Pre} & \textbf{Post} & \textbf{J} & \textbf{Pre} & \textbf{Post} & \textbf{J} \\
\midrule
\textit{CM} & 8.1 & -- & -- & 5.0 & -- & -- & 2.3 & -- & -- & 4.8 & -- & -- & 5.1 & -- & -- \\
\textit{CSh} & 9.0 & 8.3 & 41 & 8.0 & 6.0 & 11 & 8.3 & 6.0 & 10 & 7.3 & 5.3 & 6 & 11.0 & 1.3 & 6 \\
\textit{CSp} & 5.0 & 9.0 & 27 & 9.3 & 4.3 & 17 & 7.7 & 2.0 & 4 & 5.0 & 2.3 & 10 & 14.7 & 2.0 & 6 \\
\textit{CN} & 11.5 & 18.0 & 5 & 19.7 & 23.3 & 12 & 18.0 & 26.0 & 15 & 24.7 & 22.3 & 11 & 27.6 & 27.0 & 17 \\
\textit{DN} & 7.4 & 9.0 & 69 & 4.1 & 4.3 & 25 & 4.0 & 2.7 & 6 & 4.7 & 3.3 & 35 & 1.3 & 2.3 & 25 \\
\textit{EM} & 2.3 & -- & -- & 13.6 & -- & -- & 4.0 & -- & -- & 0.0 & -- & -- & 12.5 & -- & -- \\
\textit{ISN} & 8.2 & 9.3 & 72 & 6.3 & 4.3 & 12 & 7.6 & 2.7 & 19 & 3.4 & 3.3 & 12 & 1.9 & 2.0 & 22 \\
\textit{ON} & 9.0 & 9.7 & 87 & 3.7 & 4.3 & 50 & 3.3 & 2.4 & 13 & 3.7 & 3.3 & 31 & 2.3 & 1.9 & 38 \\
\textit{SeN} & 11.4 & 9.3 & 62 & 8.4 & 4.3 & 21 & 6.0 & 2.7 & 10 & 4.3 & 3.0 & 25 & 2.9 & 1.0 & 22 \\
\textit{SuN} & 7.8 & 9.3 & 52 & 5.3 & 4.3 & 35 & 3.6 & 3.0 & 20 & 3.5 & 3.0 & 29 & 1.4 & 2.0 & 25 \\
\textit{TK} & 9.3 & -- & -- & 2.1 & -- & -- & 2.8 & -- & -- & 4.9 & -- & -- & 3.8 & -- & -- \\
\midrule
\textbf{Avg.} & \textbf{8.1} & \textbf{10.2} & \textbf{45} & \textbf{7.8} & \textbf{6.9} & \textbf{18} & \textbf{6.2} & \textbf{5.9} & \textbf{12} & \textbf{6.0} & \textbf{5.8} & \textbf{15} & \textbf{7.7} & \textbf{4.9} & \textbf{15} \\
\bottomrule
\end{tabular}

\end{table*}

\subsection{Datasets}
\label{sec:datasets}

We use two primary benchmarks to evaluate and compare our metamorphic framework for testing RAG systems across different domains and reasoning requirements. 

\subsubsection{T2-RAGBench}
The financial question-answering scenario is based on T2-RAGBench~\cite{strich2025t}, a benchmark designed to evaluate RAG over large collections of financial documents containing both textual and tabular information.

T2-RAGBench includes four datasets: ConvFinQA~\cite{chen2021finqa}, FinQA~\cite{chen2021finqa}, TAT-DQA~\cite{zhu2022towards}, and VQAonBD~\cite{raja2023icdar} that cover heterogeneous financial sources such as annual reports, regulatory filings, and financial statements. All four datasets provide predefined test queries and verified answers. Test queries in these datasets typically require multi-step reasoning over retrieved evidence, such as locating specific tables or passages, extracting numerical values, and performing comparisons or arithmetic operations.

\subsubsection{RepLiQA}
We utilise the RepLiQA benchmark~\cite{monteiro2024repliqa}, which is a question-answering dataset specifically curated to benchmark LLMs on unseen reference content. It comprises \textit{synthetic} reference documents crafted by human annotators, depicting imaginary scenarios, people, and places across 17 distinct categories. Because these fictional entities and events are entirely absent from the internet, they could not have been included in the pre-training corpora of existing LLMs. Each sample in RepLiQA includes a reference document, a specific question about the document's topic, and a ground-truth answer derived directly from the text. By relying on purely synthetic content, RepLiQA guarantees that accurate answers can only be generated if the RAG system successfully retrieves the correct document and the LLM accurately comprehends the provided context, rather than relying on memorised facts.

\subsubsection{Filtering, Sampling, and Cost}
Although datasets in our evaluation were designed to preclude data contamination during LLM training, we nevertheless subject all datasets to the filtering procedure described in Section~\ref{sec:filtering}. We retain only the context dependent triplets $\langle q, \mathcal{C}, a_{gt} \rangle$. This filtering reduces the initial datasets to a pool of qualified instances: 1,868 for ConvFinQA, 3,139 for FinQA, 3,412 for TAT-DQA, 7,791 for VQAonBD, and 6,888 for RepLiQA.

However, even with the refined pools, evaluating the entire population remains economically (or computationally) infeasible to us. Each base instance yields 11 mutated executions. Applying these across two mutation modes does not strictly double the count: since three operators are inapplicable to post-chunk mode (Section~\ref{sec:operators}), we execute $11 + 8 = 19$ total mutants per triplet. Every mutant requires LLM API calls for both execution and evaluation. Our framework uses 2--3 LLM calls per mutation test: one for RAG generation on the mutated corpus, one for the semantic preservation judge, and one for the answer equivalence judge (the latter is replaced by exact numeric comparison on financial datasets, reducing the count to 2). For running RAGAS, each of its four metrics invokes the LLM independently (\textit{Faithfulness}: 2 calls for statement extraction and inference; \textit{Answer Relevancy}: 3 calls for question generation; \textit{Context Precision} and \textit{Context Recall}: 1 call each), totalling 7 LLM calls per evaluation point. For a campaign of $N$ entries and $M$ operators, we use $N{\times}M{\times}3$ LLM calls, while RAGAS uses $ N{\times}M{\times}7$. We note that this cost breakdown is not a comparative efficiency claim between ours and RAGAS, which serve different evaluation purposes; rather, it quantifies the cumulative API cost that necessitated sampling instead of exhaustive evaluation. To keep the experiment tractable while preserving statistical validity, we restrict the scope of the evaluation. From each of the five filtered datasets, we sample $n = 300$ base instances. This yields $300 \times 19 = 5{,}700$ mutants per dataset, and a total of 28{,}500 mutants across five datasets. All results are reported based on these samples and mutants.


\subsection{Configurations}
\label{sec:configurations}

\begin{table}[t]
\centering
\small
\setlength{\tabcolsep}{4pt}
\renewcommand{\arraystretch}{1.0}
\caption{\new{Mutation parameters used in the experiments.}}
\label{tab:mutation_parameters}
\begin{tabular}{@{}lll@{}}
\toprule
\new{\textbf{MO}} & \new{\textbf{Pre-chunk}} & \new{\textbf{Post-chunk}} \\
\midrule
\new{\textit{SeN}} & \new{$0.20n_s$ sentences} & \new{1 sentence/chunk} \\
\new{\textit{DN}}  & \new{$n_s/3$ fragments} & \new{$n_s/3$ fragments} \\
\new{\textit{ISN}} & \new{$0.15n_s$ sentences} & \new{1 sentence/chunk} \\
\new{\textit{CN}}  & \new{1 passage/document} & \new{1 passage/chunk} \\
\new{\textit{SuN}} & \new{$0.20n_s$ sentences} & \new{1 sentence/chunk} \\
\new{\textit{ON}}  & \new{Error rate 0.01} & \new{Error rate 0.01} \\
\new{\textit{CM}}  & \new{Merge factor 2} & \new{N/A} \\
\new{\textit{CSp}} & \new{Split factor 2} & \new{Split factor 2} \\
\new{\textit{CSh}} & \new{Offset 0.5 (128 words)} & \new{10-token shift} \\
\new{\textit{EM}}  & \new{\texttt{small} $\rightarrow$ \texttt{large}} & \new{N/A} \\
\new{\textit{TK}}  & \new{$k=1 \rightarrow 3$} & \new{N/A} \\
\bottomrule
\end{tabular}

\vspace{2pt}
\parbox{\columnwidth}{\footnotesize\new{$n_s$ denotes the number of sentences. Sentence-based quantities are rounded down with a minimum of one. EM changes between the \texttt{text-embedding-3-small} and \texttt{text-embedding-3-large} models.}}
\end{table}

We run closed-book filtering ten times, retaining only those triplets that consistently satisfy these conditions across all trials. In the study, the default top-$k$ is set to $1$.
\new{Table~\ref{tab:mutation_parameters} reports the parameter settings used in the experiment. For pre-chunk mutations, \textit{SeN} and \textit{SuN} insert sentences corresponding to 20\% of the original sentence count, \textit{ISN} uses 15\%, and \textit{DN} inserts one fragment per three sentences. At post-chunk scope, \textit{SeN}, \textit{ISN}, and \textit{SuN} insert one sentence per retrieved chunk, while \textit{DN} inserts one fragment for every three sentences. \textit{CN} adds one counterfactual passage per document or retrieved chunk, and \textit{ON} applies a character-level error probability of 0.01. The structural operators use a merge or split factor of 2, while \textit{CSh} uses a 0.5 offset before chunking and a 10-token shift after retrieval. Finally, \textit{EM} changes the embedding model from \texttt{text-embedding-3-small} to \texttt{text-embedding-3-large}, and \textit{TK} increases $k$ from 1 to 3. In the repair study, we do not evaluate $k=3$ as a repair for faults induced by \textit{TK} as it reproduces the mutated retrieval setting; we still evaluate these faults using $k=5$ and LLM-based reranking. Sentence-derived quantities are rounded down with a minimum of one.}

\new{Default RAG generation, counterfactual generation, closed-book filtering, semantic-preservation judging, and answer-equivalence judging use OpenAI’s GPT-5-mini\footnote{\texttt{gpt-5-mini-2025-08-07}}. As the GPT-5 family is composed of reasoning models, the API does not support user configurable sampling parameters such as temperature or top-$p$; we therefore use the API defaults. We use \texttt{text-embedding-3-small}\footnote{\url{https://platform.openai.com/docs/models/text-embedding-3-small}} as the embedding model.} We use a RAGAS fault threshold of $0.3$. While the artefact also includes results for thresholds of $0.0$ and $0.5$, these variations do not significantly alter the findings.
All experiments are conducted on a machine equipped with an Apple M3 Max and 36\,GB of RAM, running Tahoe 26.2.



\section{Results}
\label{sec:results}

\subsection{RQ1 (Fault Detection)}
\label{sec:results_rq1}

Table~\ref{tab:rq1_violations_overlap} reports MR violation rates and fault overlap (Jaccard index, J) by mutation operator and scope. Average violation rates range from 6.0\% to 8.1\% for pre-chunk mutations, and from 4.9\% to 10.2\% for post-chunk mutations. \textit{CN} consistently produces the highest violation rates across all datasets and both scopes, reaching 27.6\% (pre) and 27.0\% (post) on VQAonBD, showing that injecting counterfactual content directly degrades both retrieval relevance and generation faithfulness. Chunking operators \textit{CSh} and \textit{CSp} also produce elevated pre-chunk rates (e.g., 14.7\% for \textit{CSp} on VQAonBD), because altering chunk boundaries before indexing changes which content is retrieved. In contrast, surface-level operators (\textit{ON}, \textit{SuN}, \textit{DN}) yield lower violation rates overall, indicating that RAG pipelines tolerate minor formatting and typographic perturbations well.

To illustrate, consider two RepLiQA faults (i.e., True Positives). When asked ``What was a significant obstacle Emily faced during her marathon training?'', the baseline system correctly answered ``a minor ankle sprain.'' After a semantics-preserving \textit{CSh} mutation that merely shifted chunk boundaries, the system instead answered ``balancing a full-time job with her running schedule,'' retrieving a different passage fragment and producing a factually different answer despite no change to the underlying content. Conversely, when a semantics-altering \textit{CN} mutation injected counterfactual information about a yoga studio's new offering, the system was asked ``What new offering did ZenSpace Yoga launch in early 2024?'' and still answered ``virtual reality meditation classes'' verbatim, ignoring the contradictory evidence now present in the retrieved context.

The Jaccard column (J) reveals that pre-chunk and post-chunk mutations largely expose different faults. Average overlap is only 12--18\% on four of five datasets, with RepLiQA being the exception at 45\%. Surface-level operators (\textit{ON}, \textit{ISN}, \textit{DN}) show relatively high overlap because these perturbations are too minor to meaningfully alter retrieval. Both scopes end up probing the same generator-side sensitivity, and the fault sets converge. In contrast, \textit{CN} exhibits very low overlap. Overall, the low average overlap confirms that the two mutation scopes are complementary: using both would broaden fault coverage beyond what either scope achieves alone. RepLiQA's higher overlap (45\%) can be explained by its design, where each query targets one short document, so the retriever is likely to return similar chunks regardless of whether the mutation is applied before or after chunking. 

To validate the automated verdicts, two authors independently labelled 100 randomly-sampled RepLiQA mutation pairs for answer equivalence and semantic preservation. Inter-annotator agreement was 98\% ($\kappa = 0.92$) and 99\% ($\kappa = 0.96$), respectively.\footnote{$\kappa$ denotes Cohen's kappa, a chance-corrected measure of inter-rater agreement~\cite{cohen1960coefficient}.} Each annotator agreed with the LLM judge 93--95\% of the time ($\kappa = 0.68$--$0.77$). Most disagreements (7 out of 8) occurred when the LLM judged answers as equivalent but annotators deemed them different, typically involving partial matches where the mutated answer omitted details or added precision (e.g., ``late 2023'' vs.\ ``September 12, 2023''). Annotator confidence on these disputed cases averaged 2.50 out of 5, compared to 4.58 on agreed items, indicating that disagreements cluster at genuinely ambiguous boundaries rather than reflecting systematic bias.

\new{To assess whether stochastic variation in judging affects the detected violations, we repeated the final MR verdict five times for 100 stratified RepLiQA cases. 94 cases received unanimous verdicts, and the majority verdict agreed with the initial verdict in 97 cases. SATISFIED/VIOLATED flips occurred in six cases, indicating that judgement instability is limited but not absent.}

\begin{tcolorbox}[boxrule=0pt,frame hidden,sharp corners,enhanced,
borderline north={1pt}{0pt}{black},
borderline south={1pt}{0pt}{black},
boxsep=2pt,left=2pt,right=2pt,top=2.5pt,bottom=2pt]
\textbf{Answer to RQ1:} Our metamorphic framework detected faults across all five datasets (violation rates 4.9--10.2\%), with pre- and post-chunk mutations exposing distinct faults (Jaccard overlap 12--45\%), confirming both scopes are needed. A further human study validated the automated verdicts. \new{Human validation yielded 93\%--95\% agreement with the automated judge. Across repeated oracle calls, 94\% of cases received unanimous verdicts and 97\% retained the initial verdict under majority voting.}
\end{tcolorbox}

\subsection{RQ2 (Comparison)}
\label{sec:results_rq2}

\begin{table*}[t]
\centering
\small
\caption{Meta-evaluation: MR vs.\ RAGAS metrics.
Recall, Precision, and F1 computed from TP/TN/FP/FN
classifications against ground-truth correctness.}
\label{tab:mr_vs_ragas}
\resizebox{\textwidth}{!}{%
\begin{tabular}{l|ccc|ccc|ccc|ccc|ccc}\toprule
 & \multicolumn{3}{c|}{\textbf{RepLiQA}} & \multicolumn{3}{c}{\textbf{ConvFinQA}} & \multicolumn{3}{c}{\textbf{FinQA}} & \multicolumn{3}{c}{\textbf{TAT-DQA}} & \multicolumn{3}{c}{\textbf{VQAonBD}} \\
\textbf{Metric} & \textbf{Recall} & \textbf{Prec.} & \textbf{F1} & \textbf{Recall} & \textbf{Prec.} & \textbf{F1} & \textbf{Recall} & \textbf{Prec.} & \textbf{F1} & \textbf{Recall} & \textbf{Prec.} & \textbf{F1} & \textbf{Recall} & \textbf{Prec.} & \textbf{F1} \\
\midrule
Faithfulness & 0.575 & 0.379 & 0.456 & 0.373 & 0.077 & 0.128 & 0.625 & 0.085 & 0.150 & 0.605 & 0.103 & 0.176 & 0.492 & 0.086 & 0.147 \\
Answer Relevancy & 0.707 & 0.220 & 0.336 & 0.992 & 0.070 & 0.131 & 1.000 & 0.060 & 0.113 & 0.990 & 0.060 & 0.114 & 0.987 & 0.065 & 0.122 \\
Context Precision & 0.554 & 0.369 & 0.443 & 0.181 & 0.065 & 0.096 & 0.401 & 0.103 & 0.164 & 0.113 & 0.066 & 0.083 & 0.155 & 0.089 & 0.113 \\
Context Recall & 0.641 & 0.514 & 0.570 & 0.209 & 0.060 & 0.093 & 0.645 & 0.099 & 0.171 & 0.254 & 0.090 & 0.133 & 0.272 & 0.118 & 0.165 \\
\midrule
Ours & 0.894 & 0.963 & \textbf{0.927} & 1.000 & 1.000 & \textbf{1.000} & 1.000 & 1.000 & \textbf{1.000} & 1.000 & 1.000 & \textbf{1.000} & 1.000 & 1.000 & \textbf{1.000} \\
\bottomrule
\end{tabular}}
\end{table*}

Table~\ref{tab:mr_vs_ragas} compares our MR-based framework against four RAGAS metrics in a meta-evaluation against ground-truth correctness. MR achieves near-perfect F1 across all datasets: 0.927 on RepLiQA and 1.000 on others.\footnote{Note that the perfect F1 on the financial datasets is expected: their answers are single numeric values, so MR validation via exact match coincides with meta-evaluation with GT, leaving no room for classification error (see Section~\ref{sec:threats}).} In comparison, the best-performing RAGAS metric is \textit{Context Recall}, which reaches its highest F1 of 0.570 on RepLiQA. All other RAGAS metrics remain below 0.46 F1 across all datasets.

\begin{figure}[t]
     \centering
     \begin{subfigure}[b]{0.48\columnwidth}
         \centering
         \includegraphics[width=\textwidth]{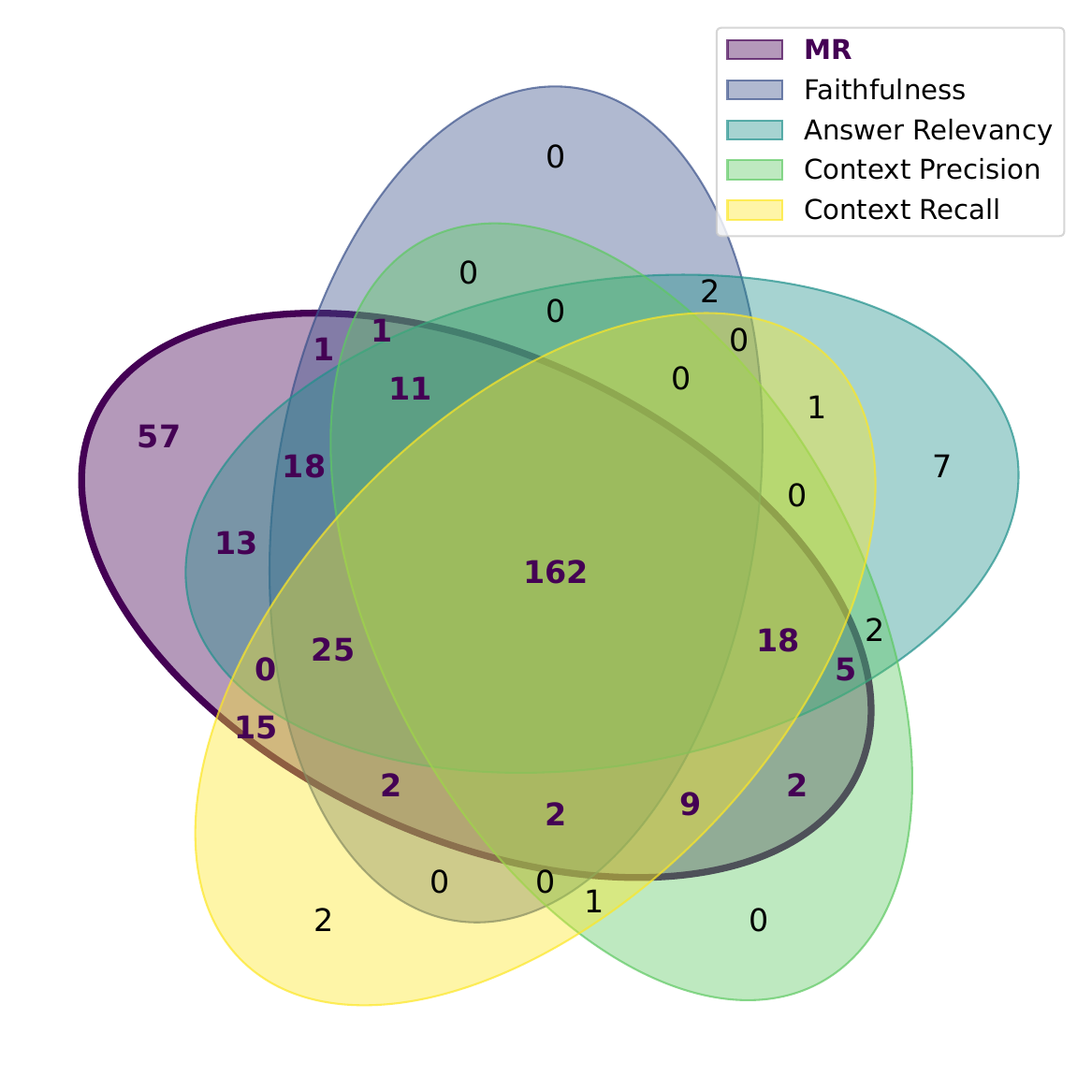}
         \caption{RepLiQA}
         \label{fig:venn_repliqa}
     \end{subfigure}
     \hfill
     \begin{subfigure}[b]{0.48\columnwidth}
         \centering
         \includegraphics[width=\textwidth]{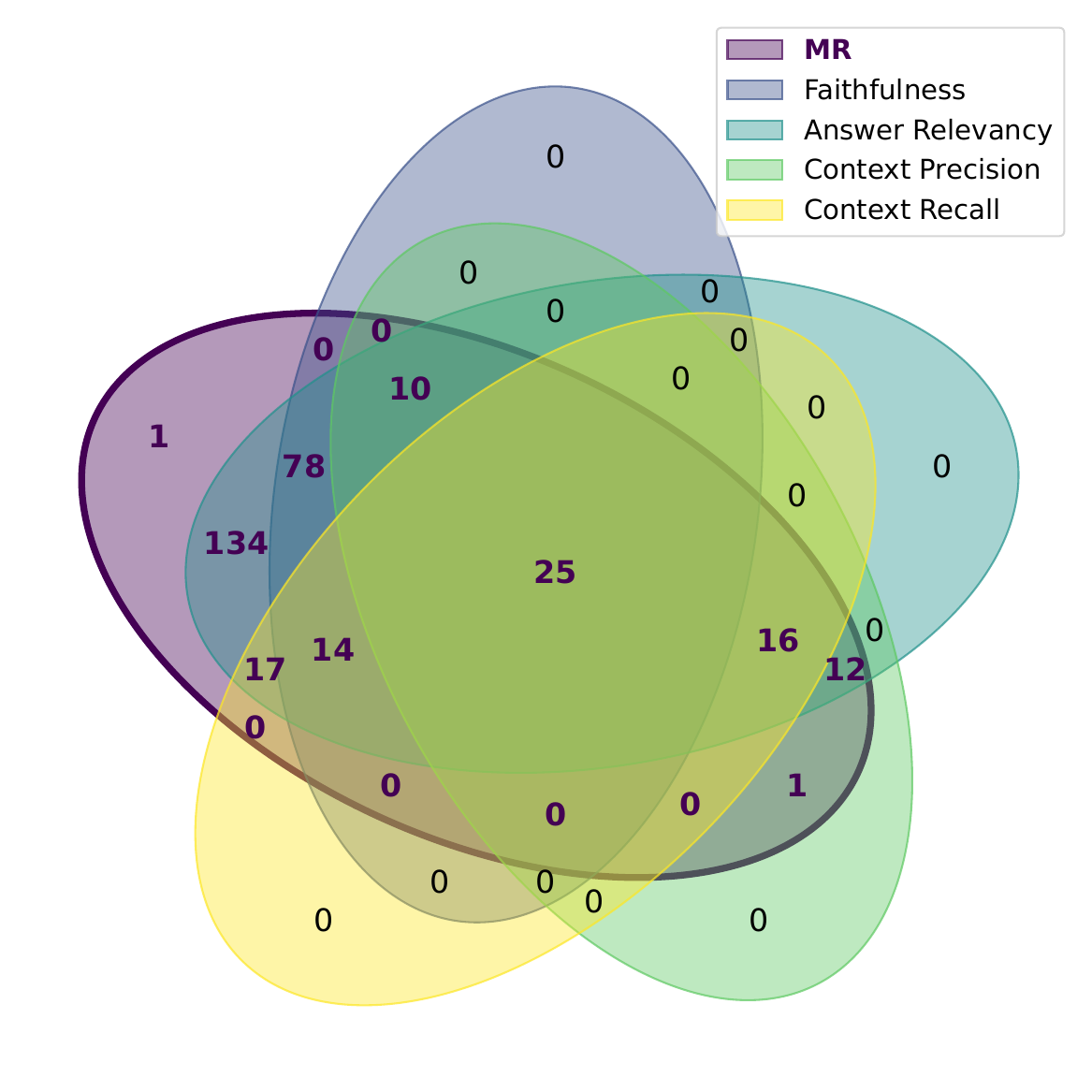}
         \caption{ConvFinQA}
         \label{fig:venn_convfinqa}
     \end{subfigure}
     \caption{Fault-overlap Venn diagrams for RepLiQA and ConvFinQA.}
     \label{fig:venn_diagrams}
\end{figure}

Figure~\ref{fig:venn_diagrams} illustrates the fault-overlap structure for RepLiQA and ConvFinQA. The remaining datasets exhibit similar patterns and are provided in our artefact (see Section~\ref{sec:data}). The diagrams depict only true-positive cases. Although metrics such as \textit{Answer Relevancy} appear to offer broad coverage, Table~\ref{tab:mr_vs_ragas} shows that all RAGAS metrics suffer from low precision and recall, so their apparent TP-only coverage overstates actual reliability. RAGAS metrics also largely overlap with each other, while MRs capture additional faults that RAGAS misses entirely.

These results demonstrate that metamorphic testing provides a more reliable diagnostic signal than score-based evaluation. By checking behavioural consistency under controlled perturbations rather than estimating quality from a single response, MRs achieve both high recall and high precision.

\begin{tcolorbox}[boxrule=0pt,frame hidden,sharp corners,enhanced,
borderline north={1pt}{0pt}{black},
borderline south={1pt}{0pt}{black},
boxsep=2pt,left=2pt,right=2pt,top=2.5pt,bottom=2pt]
\textbf{Answer to RQ2:} Our metamorphic framework substantially outperforms RAGAS metrics, achieving F1 scores of 0.927--1.000 compared to 0.083--0.570 for RAGAS. The fault-overlap analysis shows that RAGAS metrics are largely redundant with each other, whereas MRs additionally detect faults that RAGAS misses.
\end{tcolorbox}

\subsection{RQ3 (Repair)}
\label{sec:results_rq3}

\begin{table*}[t]
\centering
\small
\caption{\new{Fault repair success by dataset and scope. Each cell reports repaired/applicable faults (rate); the union counts a fault as repaired if any applicable configuration succeeds.}}
\label{tab:rq3_repair}

\begin{tabular}{l|rrr|rr|r}
\toprule
& \multicolumn{3}{c|}{\new{\textbf{Pre-chunk}}} & \multicolumn{2}{c|}{\new{\textbf{Post-chunk}}} & \new{\textbf{All}} \\
\new{\textbf{Dataset}} & \new{\textbf{$k=3$}} & \new{\textbf{$k=5$}} & \new{\textbf{Rerank}} & \new{\textbf{GPT-5.4}} & \new{\textbf{Rerank}} & \new{\textbf{Union}} \\
\midrule
\new{RepLiQA}   & \new{10/248 (4.0\%)}   & \new{11/276 (4.0\%)}   & \new{18/276 (6.5\%)}   & \new{19/243 (7.8\%)}  & \new{14/243 (5.8\%)}  & \new{53/519 (10.2\%)} \\
\new{ConvFinQA} & \new{102/212 (48.1\%)} & \new{113/218 (51.8\%)} & \new{107/218 (49.1\%)} & \new{80/166 (48.2\%)} & \new{59/166 (35.5\%)} & \new{242/384 (63.0\%)} \\
\new{FinQA}     & \new{67/196 (34.2\%)}  & \new{59/205 (28.8\%)}  & \new{62/205 (30.2\%)}  & \new{61/142 (43.0\%)} & \new{37/142 (26.1\%)} & \new{164/347 (47.3\%)} \\
\new{TAT-DQA}   & \new{78/190 (41.1\%)}  & \new{78/204 (38.2\%)}  & \new{71/204 (34.8\%)}  & \new{51/138 (37.0\%)} & \new{43/138 (31.2\%)} & \new{182/342 (53.2\%)} \\
\new{VQAonBD}   & \new{86/208 (41.3\%)}  & \new{102/215 (47.4\%)} & \new{86/215 (40.0\%)}  & \new{52/118 (44.1\%)} & \new{34/118 (28.8\%)} & \new{189/333 (56.8\%)} \\
\midrule
\new{\textbf{Overall}} & \new{343/1,054 (32.5\%)} & \new{363/1,118 (32.5\%)} & \new{344/1,118 (30.8\%)} & \new{263/807 (32.6\%)} & \new{187/807 (23.2\%)} & \new{830/1,925 (43.1\%)} \\
\bottomrule
\end{tabular}

\end{table*}

\new{Table~\ref{tab:rq3_repair} reports all repair configurations. For pre-chunk faults, direct retrieval with $k=3$ and $k=5$ repairs 32.5\% and 32.5\%, respectively, while reranking repairs 30.8\%. For post-chunk faults, upgrading the generator repairs 32.6\%, while reranking repairs 23.2\%. Altogether, the considered configurations repair 43.1\% of faults. These aggregate results mask substantial variation across datasets. Union repair rates range from 47.3\% to 63.0\% across the four financial datasets but reach only 10.2\% on RepLiQA. For pre-chunk faults, direct retrieval achieves 28.8\% to 51.8\% on the financial datasets, while both retrieval budgets repair 4.0\% on RepLiQA. Reranking similarly achieves 30.2\% to 49.1\% on the financial datasets but only 6.5\% on RepLiQA. For post-chunk faults, the generator upgrade repairs 37.0\% to 48.2\% on the financial datasets but only 7.8\% on RepLiQA, while reranking achieves 26.1\% to 35.5\% and 5.8\%, respectively. The consistently low rates on RepLiQA indicate faults that broader retrieval, stronger generation, and reranking cannot adequately address.

Repairability also varies by mutation operator. Among pre-chunk faults, \textit{EM} is the most repairable through direct retrieval with $k=5$ (49.1\%), while \textit{CSp} benefits most from reranking (49.6\%). In contrast, \textit{CN} remains difficult for both interventions, with repair rates of 16.4\% and 19.7\%, respectively. For post-chunk faults, \textit{CN} benefits most from the generator upgrade (37.3\%) but least from reranking (16.6\%), whereas \textit{SuN} achieves the highest reranking repair rate (36.9\%). These differences show that the considered interventions address distinct failure modes rather than providing a uniformly effective repair.}

As an example of a successful repair, a semantics-preserving \textit{SeN} post-chunk mutation caused the system to abstain (``I cannot answer this question based on the provided context''); upgrading the generator recovered the correct answer with full detail. Conversely, a semantics-altering CN mutation that injected a counterfactual about an app lacking currency conversion features persisted after repair: the stronger model still adopted the contradictory evidence, producing an answer that directly opposed the ground truth.

While the achieved repair rates remain partial, we note two limitations of this analysis. First, we evaluate only three general-purpose repair interventions, whereas developers may devise additional strategies based on their system and domain knowledge. Second, we do not know what proportion of the detected faults practitioners would consider worth repairing in a real deployment. Despite these limitations, the 43.1\% union repair rate indicates that the evaluated interventions can address a substantial subset of the detected faults.

\begin{tcolorbox}[boxrule=0pt,frame hidden,sharp corners,enhanced,
borderline north={1pt}{0pt}{black},
borderline south={1pt}{0pt}{black},
boxsep=2pt,left=2pt,right=2pt,top=2.5pt,bottom=2pt]
\textbf{Answer to RQ3:} \new{Repair is useful but partial. For pre-chunk faults, setting $k=3$, $k=5$, and reranking repair 32.5\%, 32.5\%, and 30.8\%, respectively. For post-chunk faults, the generator upgrade and reranking repair 32.6\% and 23.2\%. The union repairs 43.1\% of evaluated faults, but the 10.2\% repair rate on RepLiQA shows that some corpus-evolution faults persist across all interventions.}
\end{tcolorbox}


\section{Related Work}

Frameworks such as RAGAS~\cite{es2024ragas} and ARES~\cite{saadfalcon2023ares} evaluate retrieval and generation quality for individual query--answer pairs against a \textit{fixed} corpus snapshot. While effective for point-in-time benchmarking, they evaluate outputs in isolation and cannot capture whether a system behaves consistently as the corpus evolves. Our work introduces a relational, corpus-evolution-aware paradigm that surfaces failures invisible to these static metrics.

\new{MetaRAG~\cite{sok2025metarag} is closer to our setting as it also applies metamorphic testing to RAG systems, but it places the testing boundary at the response level: it mutates factoids extracted from a generated answer and checks whether the variants are supported by the retrieved context. In contrast, our system under test is the RAG pipeline under corpus evolution. Pre-chunk mutations alter documents, chunking, the index, or retrieval configuration before re-executing ingestion, retrieval, and generation, while post-chunk mutations isolate generator behaviour after retrieval. Thus, MetaRAG checks whether an existing answer is supported by its evidence, whereas our framework tests whether a RAG pipeline behaves consistently when the knowledge state changes.}

Prior work shows that LLMs are easily distracted by irrelevant retrieved passages~\cite{shi2023large}, exhibit positional bias when key evidence appears mid-context~\cite{liu2024lost}, and suppress correct parametric knowledge when retrieved content contradicts it~\cite{wu2024clasheval}. The most directly related work is Wu et al.~\cite{wu2025pandora}, which taxonomises RAG noise types empirically. Our fault taxonomy (Section~\ref{sec:fault_taxonomy}) extends theirs by adding Format Intolerance and Architectural Brittleness classes, and makes each class an executable mutation operator with an automatic oracle, rather than a descriptive analysis.

A separate line of work investigates adversarial attacks on retrieval based systems. Zhong et al.~\cite{zhong2023poisoning} show that injecting a small number of adversarial passages into a dense retrieval corpus can mislead retrieval models with high success rates, even generalising to unseen queries and domains. PoisonedRAG~\cite{zou2025poisonedrag} extends this to end-to-end RAG, achieving 90\% attack success by injecting as few as five crafted texts per target question into a knowledge base of millions. These works consider adversarial robustness, whereas our framework tests robustness under corpus evolution (reformatting, updates, noise). The two perspectives are complementary: poisoning attacks expose security vulnerabilities, while our metamorphic approach surfaces engineering faults that arise during corpus maintenance.


\section{Threats to Validity}
\label{sec:threats}

\textit{Internal validity.} \new{The LLM-based judge may misclassify semantic preservation or answer equivalence. Our human study (Section~\ref{sec:results_rq1}) found agreement between 93\%--95\%, although we acknowledge that the random sample did not cover all case categories evenly: semantics-altering cases comprised 12.0\% of the sample compared with 12.3\% of the evaluated RepLiQA population, while violated cases comprised 4.0\% compared with 9.1\%. We do not control or measure all sources of stochasticity across the complete pipeline: repeating mutation generation, e.g., the generation of counterfactuals, may produce a different execution. Our stability analysis isolates one component: we held each sampled mutation, retrieved context, and generated answer fixed and repeated only the final MR judgement five times. We observed unanimous verdicts in 94\% of cases, while the majority verdict agreed with the initial verdict in 97\%. The four financial datasets use exact numeric comparison for answer equivalence, reducing their reliance on the judge. The closed-book filter may retain queries answerable by chance; we mitigate this by running it ten times.}

\textit{External validity.} Our evaluation spans five datasets in two domains but uses a single RAG architecture and LLM. Different retrieval strategies or models may yield different fault profiles. We sample 300 base instances per dataset due to the cost of running both our framework and RAGAS across all mutation operators and scopes (28k mutants total). However, the consistent patterns observed across five independently sampled datasets provide confidence in the robustness of the findings. In future work, the cost of mutation campaigns could be reduced by prioritising the most effective operators (e.g., \textit{CN}, \textit{CSh}) based on the violation rates reported in our results.

\textit{Construct validity.} The RQ2 meta-evaluation uses ground-truth to classify faults. For RepLiQA this again relies on the LLM judge, but exact-match results on financial datasets corroborate the same findings. A potential concern is circularity in the meta-evaluation: the filtering procedure (Section~\ref{sec:experimental_procedure}) already verifies that the baseline answer $y$ matches the ground truth $a_{gt}$, so for semantics-preserving mutations the MR check ($y' \approx y$) and the GT check ($y' \approx a_{gt}$) are correlated by transitivity. We note three mitigating factors. First, the filtering is necessary for experimental validity, not to advantage MR: without it, mutations would be applied to already-incorrect baselines, making fault detection uninterpretable. Both MR and RAGAS are evaluated on the same filtered instances. Second, the correlation does not hold for semantics-altering mutations, where the GT check verifies whether $y'$ reflects the injected counterfactual, independently of $y \approx a_{gt}$. Third, RepLiQA empirically demonstrates non-trivial evaluation: MR achieves 0.927 F1, not 1.000, confirming that the LLM-based equivalence judgement ($y' \approx y$) and the GT comparison ($y' \approx a_{gt}$) can genuinely disagree when answers are natural language rather than single numeric values.


\section{Conclusion}
\label{sec:conclusion}

We presented a metamorphic testing framework for evaluating RAG systems under corpus evolution. Rather than assessing snapshot correctness, the framework defines metamorphic relations that verify behavioural consistency across 11 mutation operators applied at two scopes: pre-chunk (index-level, exercising the full pipeline) and post-chunk (context-level, isolating the generator). Across five datasets, our framework detects faults (violation rates from 4.9\% to 10.2\%) that standard RAGAS metrics miss, with human validation confirming the automated verdicts. \new{The two mutation scopes expose complementary faults. Retrieval-budget increase, generator upgrade, and LLM-based reranking together resolve 43.1\% of detected faults. While this is encouraging, their limited effectiveness on RepLiQA points to future work on alternative repair strategies.}

\section*{Data Availability}
\label{sec:data}
The implementation and the data are available.\footnote{\url{https://github.com/dbr7/RAG-metamorphic-mutation}}

\balance
\bibliographystyle{ACM-Reference-Format}
\bibliography{bib}

\end{document}